\documentclass[A4paper,12pt]{article}
\usepackage[T1]{fontenc}
\usepackage[english]{babel}
\usepackage[cp1250]{inputenc}
\usepackage{epsfig}
\usepackage{amsfonts}
\usepackage{fancyhdr}
\usepackage{syntonly}
\usepackage{graphicx}

\begin{document}
\selectlanguage{english}

\begin{titlepage}
\begin{center}
\vspace*{3cm}

\begin{title}
\bold {\Huge RHIC multiplicity distributions and superposition
models
 }
\end{title}

\vspace{2cm}

\begin{author}
\Large K. FIA{\L}KOWSKI\footnote{e-mail address:
fialkowski@th.if.uj.edu.pl}, R. WIT\footnote{e-mail address:
romuald.wit@uj.edu.pl}

\end{author}

\vspace{1cm}

{\sl M. Smoluchowski Institute of Physics\\ Jagellonian University \\

30-059 Krak{\'o}w, ul.Reymonta 4, Poland}

\vspace{2cm}

\begin{abstract}
The recent PHENIX mid-rapidity measurements of multiplicity
distributions for centrality bins are analyzed in the framework of
superposition models. A simple superposition of pp events is shown
to disagree with the heavy ion data for dispersion as a function of
centrality. However, it is suggested that a model describing better
the $pp$ data and based on the "wounded quark" idea may be
compatible with the multiplicity data for heavy ion collisions.
\end{abstract}

\end{center}

\vspace{2cm}



\end{titlepage}

\section{Introduction}

There is a long-standing inconsistency in the description of the
multiple hadron production in heavy ion collisions at high energy.
Many effects are attributed to the collective motion of quark-gluon
plasma \cite{Shu} (ideal fluid? \cite{Shu2}), or the collective
production from such a source. Thus the commonly accepted picture
adopts the idea of a collective intermediate state. However,
surprisingly large amount of data can be described by assuming the
superposition of independent nucleon-nucleon collisions as the main
mechanism of production. Therefore it is interesting to establish in
a possibly precise way the range of applicability of such an
assumption.
\par
    The multiplicity distributions in selected rapidity bin measured recently
    by PHENIX collaboration \cite{PHE} for different “centrality classes” seem to
    agree with the simple rules resulting from the superposition hypothesis, if the
    geometrical fluctuations necessarily present for each centrality class are
     subtracted from the data. However, the procedure of subtracting the
     fluctuations relies on the Monte Carlo generator which does not describe
     properly the data. Moreover, this procedure increases significantly the
     uncertainty of measurements. Therefore the lack of visible discrepancies
     with the superposition hypothesis does not prove convincingly that the
      collective effects are irrelevant for the multiplicity distributions.
 \par
    In this note we use the same PHENIX data, but do not perform any “subtractions”.
    Instead, we formulate a simple model which does not include any assumptions apart
    from the superposition idea. The heavy ion “production events” we use are simply
    the final states from the large number $N$ of superimposed $pp$ events obtained
    from the PYTHIA 8.107 generator \cite{SMS}. To each centrality class (defined by the
    range of the number of charged particles observed in the dedicated detector)
    one may estimate the range and distribution of $N$ to produce a proper sample of heavy ion events.
\par
    One should add here that such a construction does not mean
that we neglect the
    obvious effects of the screening, showering or  saturation effects summarized
    in the “wounded nucleon” \cite{BBC} models. $N$ is neither the assumed number of
    nucleon-nucleon collisions, nor twice the number of wounded nucleons.
    In fact, the estimate of the value of $N$ for the most central $AuAu$ collisions
    exceeds significantly the global number of nucleons in both colliding nuclei.
    Since we are interested just in testing the validity of the superposition
    assumption, it is enough to assume that the number $n$ of particles produced by a
    number $N_w$ of wounded nucleons is proportional to this number (with small
    fluctuations for large $N_w$). It is not necessary to assume that the proportionality
    coefficient is, e.g., half the multiplicity of $pp$ collisions at the same energy.
\par
    In the following section we give the details of our
generation procedure
    and of the definitions of quantities to be compared with data. Then we present the
    results and compare them with the PHENIX data. Short conclusions
    are contained in the last section.

\section{Assumptions and definitions}

In this note we are using the recent C++ version of the PYTHIA 8.107
generator  \cite{SMS}. We generate samples of minimum bias events
for the pp collisions at RHIC energies. To obtain the “heavy ion”
event with a selected value of $N$ we simply count all the particles
produced in a series of $N$ $pp$ events.
\par
To compare the model with the PHENIX results we have to find first
the relation between $N$ and the number $N_d$ of particles
registered in the detector BBC used to define the “centrality
class”. Thus we started by generating large numbers $N_{ev} $  of $
pp$ events divided into $N_{ev} / N$ “superevents”, each made of $N$
$pp$ events. We register then for each superevent the value of $N_d$
(counting the number of charged particles falling into the $\eta$
and $\Phi$ bins corresponding to the BBC detector) and produce the
histogram of $N_d$ corresponding to the given value of $N$. E.g.,
for the $AuAu$ collisions at 200 GeV CM energy we found the
following relations:
$$<N_d> \approx 3.9<N>,$$
 $$ D^2\approx <N_d^2>-<N_d>^2\approx 4.7 <N>$$
\par
 In principle, to produce a sample of heavy ion events corresponding to the
given range of $N_d$: $N_{min} < N_d < N_{max}$, we have  to
generate superevents for all values of N in the range for which such
values of $N_d$ can occur. Thus we generate the superevents for the
range of $N$ corresponding to an extended range $N_{min} - 3D_{min}
< N_d <N_{max} + 3D_{max}$ and remove afterwards the superevents for
which $N_d$ falls outside the required range. We have checked that
using every second, every fourth or even every eighth value of $N$
gives the same results as using all the values of $N$ in the same
range. This allows to shorten the calculations significantly. The
number of superevents generated for each value of $N$ should
correspond to the known distribution of $N_d$, which for almost all
the considered classes of centrality (except of the most central
events) falls down exponentially with a rather small coefficient in
the exponent. We assume that the distribution of $N$ has the same
shape as the measured distribution of $N_d$. The number of generated
superevents for each $N$ in the required range results from this
distribution.
\par For each of the
superevents in the sample corresponding to the given centrality
class we count the number of the charged particles $n_c$ in the
central bin of $\eta$ and $\Phi$ and $p_T$ corresponding to the
PHENIX central detector and produce a histogram of $n_c$ for this
class. As expected, the average value of $n_c$ is simply
proportional to $<N_d>$, and thus to the weighted average of $N$ in
the sample. The main non-trivial result of our analysis is the
dependence of the dispersion of $n_c$ on centrality, defined by the
range of $N_d$. This dispersion contains a contribution from the
variation of $n_c$ for given $N$ (which is simply proportional to
$N$ in all the superposition models), and a contribution reflecting
the spread of $N$ in the sample.
\par Let us repeat that we do not intend
to test any particular model, in which the dependence of average
multiplicities on energy and/or the mass of nuclei may be more or
less compatible with data. Our modeling of the heavy ion events by
superpositions of $pp$ events allows to use the distributions of $N$
as a useful data parametrization tool. Thus for each energy and
nucleus we should repeat independently the analysis of the relation
between $N$ and $N_d$ and define the proper sample of superevents to
be compared with each sample of data.
\section{Data and the superposition models}

The multiplicity distribution for the central detector (registering
the charged particles with pseudorapidity in the range $-0.26 < \eta
< 0.26$ and the range in $\Phi$ of about two units) is parametrized
by the average multiplicity <n> and the scaled dispersion squared
$$\omega = D^2/<n>.$$
If the distribution is approximated by the negative binomial
distribution (NBD) with the parameters  m and k, we have <n>=m,
$\omega = 1+m/k$. It is worth noticing that for the incoherent
superposition of K such independent sources we get the multiplicity
distribution with the average multiplicity multiplied by K, but the
same value of $\omega$.

\par
In   \cite{PHE} the authors argued that defining the centrality bin
by the range of $N_d$ one gets the NBD shape with the $k$ parameter
rescaled by a “geometrical factor” $f_{geo}$, in comparison to the
distribution at fixed (average) value of the number of nucleon
”participants”. The value of $f_{geo}$ estimated on the basis of
Monte Carlo simulations by the HIJING generator is about 0.37 for
the 200 GeV $AuAu$ data. Thus the “dynamical” value of $\omega$ is
assumed to be
$$\omega_{dyn} =
1 + f_{geo}(\omega-1)$$ and such “corrected” data are roughly
compatible for all centralities with the value measured in the  pp
collisions. Similar situation is seen for lower energies and for the
CuCu collisions. This is regarded as the argument for the absence of
collective effects in the multiplicity distributions from the heavy
ion collisions.
\par
   However, the data  show  a systematic dependence
    of $\omega$ on centrality. As we shall see, with the increasing number of participants
    there is first the increase, and then the decrease of $\omega$. The effect
    is not very strong, and becomes almost insignificant when including
    the error of the rescaling factor $f_{geo}$. Nevertheless, it seems to need an explanation.
\begin{figure}[h]
\centerline{ \epsfig{figure=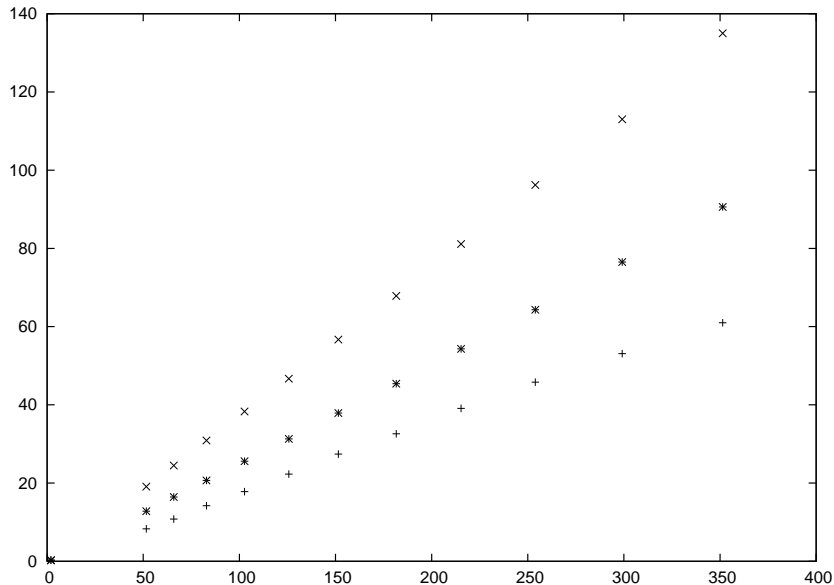,height=8.0cm}}
\caption{\footnotesize \label{Average} The average multiplicity in
the central detector for the PHENIX $pp$ and $AuAu$ data (crosses),
superposition model (stars) and the model with extended range of
$\phi$ ($x$-s) as a function of the number of participants}
\end{figure}
\par
The procedure outlined in the previous section allows to calculate
the parameters of (uncorrected) multiplicity distributions in the
central detector for various centrality cuts defined by PHENIX
experiment. The results are shown and compared with data in Figs. 1
and 2 for the $200$ GeV $AuAu$ collisions for various numbers of
participants $N_p$, as calculated in [1] for the centrality bins.
Let us note that in our model this number has no physical  meaning:
it simply labels the range of $N_d$ for each centrality class. For
comparison, the data and PYTHIA results for $pp$ collisions
($N_p=2$) are also shown.

\par
The average multiplicity, shown in Fig.1 as stars, slightly
overshoots  PHENIX data, shown as crosses (we do not show the
errors, which are of the order of the difference between the model
and data). This may be surprising, since the average multiplicity
for pp collisions calculated in PYTHIA (0.198) is much lower than
that measured by PHENIX (0.32) (note that this is not visible in
Fig.1 due to the linear scale). However, one should remember that
the centrality bins were defined by the number of particles in BBC
detectors, which for the $pp$ collisions is similarly too low in
PYTHIA. Thus to match these experimental values one needs to
superimpose a very large number of $pp$ events (reaching more than
twice the number of nucleons in two colliding nuclei!). In other
words, the correlation between the average values of $N_d$ and $N$
is roughly described in the model in which the number of
superimposed $pp$ events is chosen in such a range that the proper
range of $N_d$ is reproduced.

\begin{figure}[h]
\centerline{ \epsfig{figure=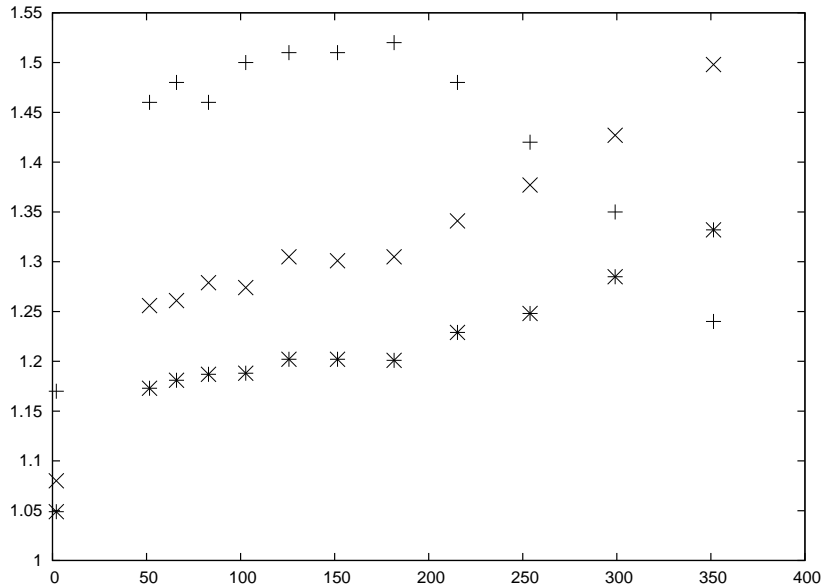,height=8.0cm}}
\caption{\footnotesize \label{Omega} The scaled dispersion for the
PHENIX $pp$ and $AuAu$ data (crosses), superposition model (stars)
and the model with extended range of $\phi$ ($x$-s) as a function of
the number of participants}
\end{figure}

\par The situation is
 different for the scaled dispersion, as shown in Fig.2. For
the peripheral and moderately central events (corresponding to the
number of participants below $200$) the values of $\omega-1$ from
PYTHIA are by a factor of $1/3$ lower than the experimental values.
Almost the same factor is found for the model and data for the $pp$
collisions. This discrepancy is not removed if one increases
artificially (by extending the range of $\phi$) the average
multiplicity from PYTHIA to fit the experimental value measured by
PHENIX for the $pp$ collisions. The resulting values, shown  as
$x$-s, are still much too low. Moreover, the agreement with data for
average multiplicity is spoiled (as shown in Fig.1).
\par Obviously,
there is some serious problem with the dispersion of the $pp$
multiplicity distribution in PYTHIA or data (or  both), and this
problem persists in the description of the $AuAu$ data for moderate
centralities. However, both in the model and in the data $\omega-1$
is approximately three times higher for the $AuAu$ than for the $pp$
data. The slow increase with centrality is also similar. Thus the
observed increase of scaled dispersion over the values from $pp$
data may be interpreted as the result of fluctuations induced by the
fluctuations in the number of participants allowed by the choice of
the range of $N_d$.
\par Certainly, it would be better to use an event
generator reproducing correctly the $pp$ multiplicity distributions
measured by PHENIX to test this interpretation, but our results
suggest that nothing more than the superposition of $pp$ events is
needed to describe the multiplicity distributions for moderate
centralities.
\par For the most central
events the disagreement is much more spectacular. The scaled
dispersion calculated for the superposition of $pp$ PYTHIA events
grows approximately linearly with $N_p$ (corresponding to the linear
increase with average $N_d$), whereas the data show the significant
decrease and for the most central bin even fall below the model
results. Thus the na\"\i ve superposition model fails to describe
the data for scaled dispersion in the central collisions. In the
next section we discuss the possible interpretation of this effect.
\par
The observed increase of $\omega$ with centrality in the
superposition model casts some doubts on the correction procedure
for the "geometrical" fluctuations applied in \cite{PHE}. In this
procedure it was assumed explicitly that a single rescaling
parameter is sufficient to correct the data for all centralities. We
see that the fluctuations due to the non-zero range of centralities
in each bin do not scale but increase significantly with average
centrality in the bin.

\section{Discussion of the results}
 In any superposition model the central collisions, with
    large number of participants, are expected to yield naturally larger multiplicity
    fluctuations than the less central ones. Is then the observed decrease of
    scaled dispersion a signal of some collective effects, or can one understand
    it within the superposition models?
\par    To answer this question, let us first remind that the values of $N$ needed to
    describe the central collisions are much higher than number of nucleons in
    the $Au$ nucleus. Thus the simple “wounded nucleon” model cannot describe the
    data: even if all the nucleons in both nuclei are wounded, the average
    multiplicity from them $<n>_{AA}$ should be just $A <n>_{ NN}$. This fact, known since
    quite a long time, gave rise to the so-called “wounded quark” model \cite{BCF}.
\par     In such a model the pp collision results in “wounding” just a single
    quark from each nucleon, but in the multiple nucleon interactions during
    a heavy ion collision more than one quark of this nucleon is usually wounded,
    resulting in the enhancement of average multiplicity. In a simple version
    of wounded quark model considered recently \cite{BB} a nucleon consists of a
    quark and a diquark, both interacting similarly and yielding similar number of hadrons when wounded.
 \par   Now let us consider the multiplicity distribution from a single nucleon
    in such a picture assuming that the distribution of products from one
    wounded quark (or diquark) may be approximated by NBD with parameters $<n>_q$ and $k_q$,
    yielding $\omega_q=1+<n>_q/k_q$. During the heavy ion collision the multiplicity distribution
    from any nucleon may be thus parametrized as a superposition of two distributions:
    from one quark (with probability $\alpha$) and from both constituents (with probability $1-\alpha$).
    It is straightforward to prove that the parameters of the resulting distribution are
$$<n>_1=\alpha<n>_q+2(1-\alpha)<n>_q=(2-\alpha)<n>_q,$$
$$ \omega_1=1+<n>_q/k_q+<n>_q\alpha(1-\alpha)/(2-\alpha).$$
 We see that both for $\alpha=1$ (only one quark wounded)
and for $\alpha=0$ (both constituents wounded) the scaled dispersion
is the same, but for the intermediate values of $\alpha$ an
additional positive term appears and $\omega_1$ has a maximum at
$\alpha=0.5$.
\par     If we consider a class of events
corresponding for some range of
    centralities (defined, e.g., by the number of participant nucleons $N$)
    and assume fixed $\alpha$ in this range,  the multiplicity distribution parameters will  read
$$<n>=<n>_1<N> $$ $$\omega = \omega_1+<n>_1\omega_N$$ where $\omega_N=(<N^2>-<N>^2)/<N>$ for the given
range of $N$. If $\alpha$ increases from  $0$ to $1$ for increasing
$<N>$, the first term passes through a maximum at $\alpha =0.5$, and
a similar maximum may appear in the dependence of $\omega$ on $<N>$.
Obviously, fixed $\alpha$ for each range of $N$ is not a realistic
assumption, but may serve as a first approximation.
\par    This suggests that the wounded quark
model may explain, at least qualitatively, the
    non-monotonic dependence of $\omega$ on centrality, as the increase of $\alpha$ with $<N>$ is a very
    natural feature of this model. For the peripheral collisions most of the nucleons
    interact only once and thus only one quark in each of them is wounded. For the
    central collisions almost all nucleons interact more than once and both of their constituents are wounded.

\section{Conclusions}

We have investigated the PHENIX data of the multiplicity
distributions in the central rapidity region for changing centrality
in heavy ion collisions. We use a simple model, in which each final
state for a heavy ion collision is constructed as a superposition of
many $pp$ events and the combination of such states is arranged to
fit the experimental definition of various centrality classes.
\par
We show that the model which describes roughly the average
multiplicity fails to describe the data on scaled dispersion. For
moderate centralities the disagreement seems to result simply from
the imperfection of the model for $pp$ collisions. For most central
events the discrepancy between the model and data is more severe,
but it is qualitatively similar to the effect expected in the
wounded quark model. A construction of a superposition model for the
multiplicity distributions for different centralities would be
desirable. Some time ago a "{}quark participants"{} model was
already shown to describe the centrality dependence of the average
multiplicity \cite{EV}.

\section{Acknowledgments} We are
grateful to A. Kotański for reading the manuscript. One of us (RW)
is grateful for a partial support from the COCOS
(MTKD-CT-2004517186) project.

\end{document}